\documentclass[aps,pre,floatfix,showpacs,preprint]{revtex4}
\usepackage{graphicx}

\usepackage[greek,english]{babel}
\begin{document}
\latintext
\title{\bf A Model of the Quantum-Classical and Mind-Brain Connections, 
   and of the Role of The Quantum Zeno Effect in the Physical Implementation
   of Conscious Intent.}
\author{Henry P. Stapp}
\affiliation{Theoretical Physics Group\\
                               Lawrence Berkeley National Laboratory\\
                                         University of California\\
                                     Berkeley, California 94720}
\date{\today}
\begin{abstract}
A simple exactly solvable model is given of the dynamical coupling 
between a person's classically described perceptions and that person's
quantum mechanically described brain. The model is based jointly upon von 
Neumann's  theory of measurements and the empirical findings of close 
connections between conscious intentions and synchronous oscillations 
in well separated parts of the brain. A quantum-Zeno-effect-based
mechanism is described that allows conscious intentions to influence brain activity
in a functionally appropriate way. The robustness of
this mechanism in the face of environmental decoherence effects is emphasized. 

\end{abstract}
\maketitle
\section{Introduction}
The basic problem in the interpretation of quantum mechanics is to 
reconcile the fact that our observations are describable in terms
of the concepts of classical (i.e., ninteenth century) physics, 
whereas the atoms from which our measuring devices and our physical 
body/brains are made obey the laws of quantum (twentieth century) 
physics. The direct application of the microscopic atomic laws to macroscopic 
aggregates of atoms is well defined, but the thus-defined aggregates 
of atoms are not describable in classical terms.
 
The basic problem of the philosophy of mind, and indeed of all philosophy,
is to understand the connection of our conscious thoughts to the physically 
described world. No feature, configuration, or activity of the physical world, 
as it is conceived of and described in classical physics, {\it is} the 
experiential quality that characterizes our conscious thoughts, ideas, and 
feelings. Something beyond the classically conceived physical world seems 
to be needed in the full inventory of what exists. 

The obvious solution to this second problem is to recognize that the basic precepts
of classical physics were replaced during the first part of the twentieth
century by those of quantum theory. This improved physical theory brings 
conscious human observer/agents into 
physics in an essential way that renders the classical conceptions of our
bodies, including our brains, fundamentally deficient. The new  theory 
accommodates a mechanism that allows our conscious thoughts to influence our 
bodily actions without being reducible to any physically describable feature
 or activity. Hence accepting the precepts of contemporary physics provides an
adequate and suitable basis of a rational answer to the second question. 
But it leaves untouched the first question, about the basis within a 
quantum universe of the classical describability of our perceptions.

Mounting neuroscientific evidence indicates that our conscious intentions 
are closely linked to synchronous $\sim 40 Hz$ oscillations of the
electromagnetic field at many well-separated brain sites\cite{fell,engel}. 
This result points to the
importance of classically describable simple harmonic oscillator (SHO) motions
in the description and understanding of our conscious intentions and their physical effects. 
But a focus on SHO motions opens the door to a relatively simple solution of the problem of the
connection between 
the classical character of our descriptions of our perceptions and the quantum
character of our description of the physical dynamics. The so-called ``coherent states'' 
associated with SHO motions connect quantum concepts to classical concepts in just the
way needed to achieve a simple, rational, simultaneous solution of the problems of the 
quantum-classical and mind-brain connections. The purpose of this article is to describe 
an exactly solvable model that exhibits in a clear way the basic elements of this 
resolution of these two problems.

In this model the causal effectiveness of our conscious intentions rests heavily  
upon the quantum Zeno effect. This is a strictly quantum mechanical  effect that has
been advanced elsewhere\cite{stapp1,stapp2,stapp3,stapp4,stapp5}   as the  
dynamical feature  that permits ``free choices'' on  the part of  an observer to  
influence his or her bodily behavior. 

The intervention by the observer into 
physical brain dynamics is  an essential  feature of  orthodox (von Neumann)
 quantum mechanics.  Within the  von Neumann quantum dynamical framework 
 this intervention  can, with  the  aid  of quantum  Zeno  effect, cause  a
 person's brain  to behave  in a way  that causes  the body to  act in
 accord with the person's  conscious intent. However, the previous accounts of
this mechanism, although strictly based on the mathematical principles of quantum mechanics,
have been directed primarily at neuroscientists and philosophers, and have therefore been largely
stripped of equations. The present model is so simple that the equations and their meanings can be
presented in a way that should be understandable both to physicists and to sufficiently 
interested nonphysicists who are not troubled by simple equations.

On the other hand, the use of quantum mechanical effects in brain dynamics might  seem
 problematic,  because   it  depends  on  the  existence  of  a
 macroscopic quantum effect  in a warm, wet, noisy,  brain. It has
 been  argued that {\it some} such effects  will be  destroyed  by environmental
 decoherence\cite{decoherence}.  Those  arguments 
 do cover many macroscopic quantum mechanical effects, but they fail, for the 
reasons described below, to  upset  the quantum Zeno effect at work here.

In section II I shall review the well known properties of a system of two coupled SHOs.
In section III I shall use those results, and the closely related properties of
the associated quantum ``coherent states'', to construct a mathematically solvable
quantum mechanical model of the connection between conscious intent and brain activity. 
In section IV I describe the conclusions to be drawn.

\section{Coupled Oscillators in Classical Physics}

It  is   becoming  increasingly   clear  that at least some of our   normal  conscious
experiences are  associated with  $\sim 40 Hz$ synchronous oscillations  of the
electromagnetic fields  at a collection of brain sites\cite{fell,engel}. These  sites 
are evidently dynamically coupled.  
And the brain appears to be, in some sense, approximately described by classical physics. So I begin
by  recalling some  elementary  facts about  coupled classical  simple
harmonic oscillators (SHOs).

In suitable units  the Hamiltonian for two SHOs  of the same frequency
is
\begin{eqnarray}
H_0 ={1 \over 2} (p^2_1 + q^2_1 + p^2_2 + q^2_2).
\end{eqnarray}
If we introduce new variables via the canonical transformation
\begin{eqnarray} 
P_1 &=& {1 \over {\sqrt{2}}} (p_1 + q_2)\\
Q_1 &=& {1 \over {\sqrt{2}}} (q_1 - p_2)\\
P_2 &=& {1 \over {\sqrt{2}}} (p_2 + q_1)\\
Q_2 &=& {1 \over {\sqrt{2}}} (q_2 - p_1),
\end{eqnarray}
and replace the above $H_0$ by
\begin{eqnarray}
H = (1+e)( P^2_1 + Q^2_1)/2 + (1-e)( P^2_2 + Q^2_2)/2,
\end{eqnarray}
then this $H$ expressed in the original variables is
\begin{eqnarray}
H = H_0 + e(p_1 q_2 - q_1 p_2).
\end{eqnarray}

If  $e <<1$ then  the term  proportional to  e acts  as a  weak coupling
between the two SHOs whose motions for $e = 0$ would be specified by $H_0$.

The Poisson  bracket (classical) equations  of motion for  the coupled
system are, for any $x$,
\begin{eqnarray}
dx/dt &=& \{ x, H \} = \sum_j \Bigl ( {{\partial x} \over {\partial q_j}} 
{{\partial H} \over {\partial  p_j}} - {{\partial x} \over {\partial p_j}} 
{{\partial H} \over {\partial q_j}}\Bigr ).
\end{eqnarray}
They give
\begin{eqnarray}
dp_1/dt &=& - q_1 + p_2 e\\
dp_2/dt &=& - q_2 - p_1 e\\
dq_1/dt &=& p_1 + q_2 e \\
dq_2/dt &=& p_2 - q_1 e.
\end{eqnarray}

A solution is
\begin{eqnarray}
p_1 &=& {C \over 2} [ \cos (1+e)t + \cos (1-e)t ] \nonumber\\
    &=&  C \cos t \cos et   \\ 
q_2 &=& {C \over 2} [ \cos (1+e)t - \cos (1-e)t ]  \nonumber\\
    &=& -C \sin t \sin et   \\
p_2 &=& {C \over 2} [ - \sin (1+e)t + \sin (1-e)t ] \nonumber\\ 
    &=& -C \cos t \sin et   \\
q_1 &=& {C \over 2} [ \sin (1+e)t +  \sin (1-e)t ]  \nonumber\\
    &=& C \sin t \cos et  .
\end{eqnarray}
The second line of each equation follows from the trigonometric formulas for sines 
and cosines of sums and differences of their argumants. A common phase $\phi$ can be 
added to the argument of every sine and cosine in the first line of each of the
four equations. This leads to the addition of this phase to the argument t, but not 
the argument et, in the second line of each of the four equations. 

These equations  specify the  evolving state of  the 
two SHO system  by a trajectory  in $(p_1,  q_1, p_2,  q_2 )$  space.

When  we introduce  the  quantum corrections  by quantizing  this
classical model we obtain  an almost identical quantum mechanical
description  of the  dynamics. In  a very well  known way  the
Hamiltonian $H_0$ goes over to  (I use units where Planck's constant
is $2\pi$.)
\begin{eqnarray}
  H_0 &=& {1 \over 2} (p^2_1 + q^2_1 + p^2_2 + q^2_2) =
(a^{\dagger}_1 a_1 + 1/2) + (a^{\dagger}_2 a_2 + 1/2).
\end{eqnarray}
The connection  between the classical and quantum  descriptions of the
state of  the system is very  simple: the point  in $(p_1, q_1, p_2, q_2)$
space  that represents  the classical  state  of the  whole system  is
replaced  by  a  ``wave  packet''  that, insofar  as  the  interventions
associated  with  observations can  be  neglected,  is  a smeared  out
(Gaussian) structure centered for all  times exactly on the point that
specifies  the classical  state of  the system.  That is,  the quantum
mechanical  representation  of   the  state  specifies  a  probability
distribution of the  form ($exp(-d^2)$  ) where $d$ is the  distance from a
center (of-the-wave-packet) point $(p_1, q_1,  p_2, q_2$ ), which is, at all
times,  exactly the  point $(p_1,  q_1, p_2,  q_2 )$  that is  the classical
representation of the state.

According to  quantum theory, the operator  $a^{\dagger}_i a_i = N_i$  
is the number
operator that gives  the number of quanta of type $i$  in the state.

Thus in the absence of any observations the classical and  quantum descriptions 
are  almost identical: there is, in the quantum treatment, merely a small 
smearing-out in $(p,q)$-space, which is needed to satisfy the uncertainty 
principle.

This  correspondence  persists  when  the coupling  is  included.  The
coupling term in the Hamiltonian is
\begin{eqnarray}
H_1 &=& e (p_1 q_2 - q_1 p_2 - p_2 q_1 + q_2 p_1)/2 \nonumber \\
   &=& ie/2 (a^{\dagger}_1 a_2 - a_1 a^{\dagger}_2 - a^{\dagger}_2a_1 
+ a_2 a^{\dagger}_1) .
\end{eqnarray}
The  Heisenberg  (commutator) equations  of  motion  generated by  the
quadratic Hamiltonian $H  = H_0+ H_1$ gives the  same equations as before,
but now with operators in  place of numbers. Consequently, the centers
of the wave packets will follow the classical trajectories also in the
$e > 0$ case. The radius of the orbit is the square root of twice the 
energy, measured in the units defined by the quanta of energy associated 
with frequency of the SHO. 

\section{Application}

With these well known results in hand,  we can
turn to their application.  The above mathematics shows, for SHOs, a
near identity between the classical and quantum treatments, insofar as 
there are no observations. But if observations occur, then the quantum dynamics
prescribes certain associated actions on the quantum state. 

The essential point here is that quantum theory, in the von Neumann/Heisenberg 
formulation, describes the dynamical connection between conscious observations 
and brain dynamics. [Von Neumann\cite{von-neumann} brought the mind-brain connection into
the formulation in a clear way, as an application of the orthodox quantum precept
that each increment in our classically describable knowledge is represented in the 
mathamatical language of quantum mechanics by the action of associated  projection 
operators 
on the prior state. Heisenberg\cite{heisenberg} emphasized that if one wants
to understand what is really happening then the quantum state should be regarded 
as a ``potentia'' (objective tendency) for a real psycho-physical event to 
occur.] To apply this theory, the classically described brain must
first be converted to its quantum form. By virtue of the relationships described
in section II, this conversion is direct when
the classical state that is connected to consciousness is a SHO state.
And if no observations occur the classical and quantum descriptions are essentially
the same: the tiny smearing out of the classical point to the narrow gaussian centered 
on the classical point is of negligible significance.  

The observer, in order  to get information about what  is going  on about him  
into his stream  of consciousness, must, according to orthodox quantum mechanics,
initiate  probing actions. According  to the development of the theory
of von Neumann\cite{von-neumann}  described in
Refs\cite{stapp1,stapp2,stapp3,stapp4,stapp5}, the brain  does  most of  the 
work.  It creates, in  an essentially mechanical  way based on trial and 
error learning, and also upon the current quantum state of the brain, a 
query/question. 
Each possible query is  associated with a psychological projection into the
future that specifies the  brain's computed ``expectation''  about what the 
feedback from the  query will be. [My idea here is to assume/postulate that if 
${P1, P2, P3, ...}$
is the set of $P's$ corresponding to all the questions that could be posed at 
time $t$, and $P(t)$ is the $PN$ that maximizes $Trace \; PN \rho (t)$, 
then the only question that could 
be asked at time $t$ is $P(t)$. But whether this question will in fact be
posed at time $t$ could be influenced by experiential qualities. This would 
allow the {\it timings} of the probing actions to be determined in part by
features of nature not represented in the physically described part.]

The  physical
manifestation of this query is called  ``process 1'' by von Neumann. It is
a  key and necessary element of the quantum dynamics: it resolves ambiguities
that are not resolved by the physical laws of quantum mechanics,
and it ties the physical description expressed in terms of the quantum 
mathematics to our communicable descriptions of our perceptions. This process 
1 probing action is {\it not} the famous statistical element in quantum theory! 
It is needed both in order to specify what the statistical predictions will be
{\it about}, and also to tie the abstract quantum mathematics to human 
perceptual experience, and hence to science.

In order to bring out the essential point,  and also to tie  the discussion
comfortably  into  the common understandings  of  neuroscientists,  who  are
accustomed to  thinking that the brain  is well described  in terms of
the  concepts  of  classical   physics,  I  shall  consider  first  an
approximation  in  which the  brain  is  well  described by  classical
ideas. Thus the two SHO states  that we are focusing on are considered
to be aspects of possible states of a classically described brain, 
which is also
providing the potential wells  in which these  two SHOs move.  It is 
the  degrees of freedom of the brain associated with the first of these two SHOs 
that are, in the simple model being considered  here, the possible 
brain  correlates of the  consciousness of
the observer  during the  period of the  experiment. Hence it  is they
that are affected by von Neumann's process 1. The second SHO, described
by the pair of variables $(p_2, q_2)$, represents environmental degrees
of freedom. One sees from the second lines in each of the four equations (13-16) that 
in a period of duration $t=\pi (2e)^{-1}$ starting from time $t=0$   
the energy of the first SHO will, for $e<<1$, be fully transferred to 
the second SHO, provided no probing actions are made. 

If no probing actions are made then the conserved energy will oscillate
with period $t=2\pi (e)^{-1}$ back and forth between the two SHOs.
Our interest here is in the effect upon this transfer of energy from the first 
SHO  to the second SHO of a sufficiently rapid sequence of probing actions. 
What will be shown is that if the probing actions are sufficiently rapid on the scale
of time $t= e^{-1}$ then the trajectory of $(p_1, q_1)$ will tend to follow the uncoupled
$(e=0)$ trajectory.

The point, here, is
that quantum mechanics has a built-in connection between a conscious intent and its
physical effects. This connection is tied to the process 1 probing actions,
whose dynamical effects are specified by the quantum dynamical rules. Therefore
our conscious intentions do not stand outside the dynamics as helpless, impotent
witnesses, as they do in classical physics, but have {\it specified} dynamical effects. 
We are now in a position to examine what these effects are.

I assume that there is a rapid sequence of queries at a sequence of times
$\{t_1, t_2, t_3, ...\}$. These queries will be based
on expectations constructed by the brain on the basis of past experiences. 
These queries are represented in the quantum mathematics by a series of 
projection operators $\{P(t_1), P(t_2), P(t_3), ...\}$ 
[A {\it projection} operator $P$ satisfies $PP=P$] This sequence of projection operators 
represents a sequence of questions that ask whether the current state is
on the ``expected'' track. This track is specified by the $e = 0$ trajectory,
which represents expectations based on past experiences in which the 
holding-in-place effects of similar efforts have been present. 

Up until now I have spoken as if the projection operators associated with
the observations are projections onto a single quasi-classical state (i.e.,
onto one of the so-called ``coherent' states.) 
A projection upon such a state 
would involve fantastic precision. Each such state is effectively 
confined to a disc of unit size relative to an orbit radius $C$ of 
about $10^{6}$ in the units employed in equation (1). [This number $10^{6}$
is roughly the square root of the thermodynamic
energy per degree of freedom at body temperature, in energy units associated
with equation (1), in which Planck's constant is 2$\pi$, and the angular velocity is
one radian per unit of time. The unit of time in these units is about 4ms for
40 Hertz oscillations. An actual excited brain state should have energy significantly
{\it greater} than thermal, but a higher energy makes our approximation even better].
However, it it is possible (for our SHO case) to 
define more general operators that are projection operators 
(i.e., satisfy $PP = P$) apart from corrections of 
order, say, $ <10^{-3} $, by using the von Neumann lattice 
theorem.\cite{klauder} 

If one represents by $[P,Q]$ the projection operator that projects onto the 
Gaussian state centered at $(p, q)=(P,Q) $, 
then the lattice theorem says that the following identity holds:
\begin{eqnarray}
 \sum [mf,nf] = I
\end{eqnarray}
where $ f=(2\pi )^{1/2} $ , $I$ is the identity operator, and 
the sum is over 
all integer values 
of $m$ and $n$ except $ m = n = 0 $.
Moreover, the decomposition into different Gaussian components effected 
by this identity is unique. If one restricts the sum to the lattice points 
in a very large square region in $(p_1, q_1)$ space then the resulting 
operator $P' $ is very nearly a projection operator. 

For example, if the square region $S(C, 0)$ is centered
at the SHO point $(C,0)$ in the $(p_1, q_1) $ space that we have been discussing,
and has sides of length, say, one percent of the radius $C $ of the unperturbed orbit,
then each side of the square will be $10^{4} f^{-1} $ units compared to the unit
size associated with the Gaussian fall off, $\exp (-d^2)$. In this case the associated 
quasi-projection operator $P'=P(C,0)$ is essentially a projection operator
onto the square region $S(C,0)$ of $(p_1,q_1) $ space: it will take any state
vector, uniquely decomposed into the sum of terms specified in equation (19), 
approximately into the sub-sum over the terms occurring in $P'$.

Let $S(C\cos \phi , C\sin \phi)$ be the square, centered on $(C\cos \phi ,C\sin \phi)$,
obtained by rotating $S(C,0)$ by $ \phi $, so that the line from its center point to 
the origin is parallel to two of its sides. The action of the unperturbed $(e=0)$ 
Hamiltonian will take $S(C,0)$ to $S(C\cos \phi ,C \sin \phi) $ in time $\phi $.
It will also take $P(C, 0)$ to the quasi-projection operator 
$P(C\cos \phi ,C\sin \phi)$ associated with the
square $S(C\cos \phi , C\sin \phi)$. These results follow from the simple SHO
dynamics in the unperturbed (decoupled) $e=0$ case.

The collapse rules of orthodox quantum dynamics are compactly stated in terms 
of the $Trace$ operation.
[The $Trace$ operation acting upon operators/matrices is defined by allowing 
the matrix (or operator) multiplication operation occurring in, say, $Trace AB$ to 
be extended cyclically, so that $B$ acting to the right acts back on A. This 
means that for any pair of matrices/operators $A$ and $B$, $Trace \; AB = 
Trace \; BA$. 
This property entails also that $Trace \; ABC =Trace \; BCA$. For any $X$, 
$Trace \; X$ is a number. In our case, $Trace \; P(P,Q)$ is essentially the area
of the square S(P, Q), measured in units of action given by Planck's
constant, and $Trace \; P(P,Q) P(P', Q')$ is the area of the intersection of
$S(P,Q)$ and $S(P', Q')$.]
  
The $Trace$ of the product of the ``projection'' operator $P(p_1(t,e),q_1(t,e))$ centered 
on the perturbed orbit [where the two arguments are defined by equations (13) and (16)] 
with the ``projection'' operator $P(p_1(t,0),q_1(t,0)) = P(C\cos t, C\sin t)$ 
centered on the unperturbed orbit is, to lowest order in $t $, 
$1 - {1 \over 2 } ((et)^2  100) $, 
where for a 40 Hertz SHO the time unit is about 4 ms. The term 
${1 \over 2} ((et)^2  100) $ is the ratio of the displacement [of the perturbed
square relative to the unperturbed square, namely ${C \over 2} (et)^2$],  to the 
length of the side of the square, 
which is one percent of the radius $C$ of the unperturbed orbit. The unperturbed
square rotates rigidly with angular velocity unity, under the action of the 
unperturbed Hamiltonian, and the lowest-order $e>0$ displacement is toward the origin
$(p_1, q_1) = (0, 0)$. Consequently, the dynamics is essentially unchanged
by  rotations: the initial condition $(C,0)$ plays no essential role. 

According  to  the  basic  precepts  of  quantum  theory, 
the (physical) ``state'' of the system at time $t$ is specified by a 
``density matrix'' (or ``density operator''), usually denoted by 
$\rho (t)$. If the answer is 'Yes', then the state immediately {\it after} the 
probing action at time $t_i$ is $\rho (t_i+) = P(t_i)\rho (t_i-) P(t_i)$, 
where $\rho (t_i-) $ is the state immediately 
{\it before} the time $t_i$ at which the question is posed. The operators $P(t_i)$
that occur on the right and left in $\rho (t_i+)$ project onto states that 
in our case are evolving at time $t_i$ according to the unperturbed $(e=0)$ SHO motion.
Hence for our case the first-order evolution 
forward in time from the probing time $t_i$ is the same as the {\it unperturbed}
$(e=0)$ evolution. This means that the small-time evolution forward in time by the time
interval $t$ from the time $t_i$ of the $i^{th}$ probing action is given by the 
second lines of equations (13) and (16) with the arguments $t$ in those
two equations replaced by $t+t_i$ but the arguments $et$ left unchanged.

The basic statistical law of quantum theory asserts that, {\it given the 
query specified by the projection operator $P(t)$ }, the probability that the 
answer will be 'Yes' is $Trace \; \rho (t+)$ divided by $Trace \;  \rho(t-)$.

Note that the query, specified by $P(t)$ and by the time $t$ at which $P(t)$ acts,
must be specified {\it before} the statistical postulate can be applied!
 
If $\rho (t-)$ is, for the first probing time $t=t_1$, slowly varying over the
square domain in $(p_1,q_1)$ space, 
in the sense that $Trace \;  [P,Q]\rho (t-)$ is essentially constant as $(P,Q)$ varies
over the square $S(C, 0)$, 
then the state immediately after the initial observation will be essentially the projection 
operator $P(C, 0)$ associated with that initial process 1 probing action.  

Under these conditions our equations 
show that for any (large) time $T$ the density matrix $\rho (T)$ will be nearly equal to 
$P(C\cos T,C\sin T)$, provided the interval $T$ is divided by observations 
into $N$
equal intervals $t_{i+1} - t_i$, and $N (10eT)^2 N^{-2} <<1$. This condition
entails both that {\it all} the answers will be 'Yes' with probability close to unity, 
and also that the final $\rho (T)$ will be almost the same as the unperturbed 
``projection'' operator $P(C\cos T, C\sin T)$. 

Thus the rapid sequence of probing actions effectively holds the sequence of outcomes to the
{\it expected} sequence. The affected brain states are constrained to follow the {\it expected}
trajectory! This is the quantum Zeno effect, in this context.

This result means that if the probing actions come repetitiously at
sufficiently short time intervals  then the probability that the state 
will remain on the unperturbed orbit for, say, a full second will remain  
high even though the perturbed $e>0$ classical trajectory moves away from the 
unperturbed orbit by an  amount of order $C$ in time $T$ of order $e^{-1}$.

The drastic slowing of  the divergence of  the actual orbit from the 
computed/expected (circular-in-this-case)  orbit is a manifestation of
the quantum Zeno effect. The representation in the physically described
brain of the probing 
action corresponding to the query ``Is the brain  correlate of the
occurring  percept  the  computed/expected state''  is von  Neumann's
famous process  1, which  lies at the mathematical  core of  von Neumann's
quantum  theory  of  the  relationship between  perception  and  brain
dynamics.

\section{Conclusions}

The bottom line is that orthodox quantum mechanics has a built-in 
dynamical connection between conscious intent and its physically describable
consequences. This connection fills a dynamical gap in the purely physically
described quantum dynamical laws, and it allows certain specific mind-brain connection to
be {\it deduced from the  basic physics precepts relating mind and brain.}
If a person can, by 
mental effort, {\it sufficiently increase the rate at which his process 1 probing 
actions occur} [this is something not under the control, even statistically, of 
the physical laws of quantum mechanics] then that person can, by mental effort, 
quantum  dynamically {\it cause}
his brain/body to behave in a way that follows a pre-programmed trajectory, specified,
say, by ``expectations'', instead of following the trajectory that it would follow
if the von Neumann process 1 probing actions do not occur in rapid succession.
Because the causal origin of the process 1 probing actions {\it is not specified,
even statistically, by the presently known laws of physics,} there is in quantum 
mechanics a rational place for the experiential
aspects of our description of nature to enter, irreducibly and efficaciously, into the 
determination of the course of certain physically described events.

I have focused here on the leading powers in t, in order to emphasize,
and exhibit in  a relatively simple way, the origin of the
key result, which is that for small t on the scale, not of the 
exceedingly short period  of the  quantum mechanical  oscillations, 
nor  even on the $\sim 25 ms$
period of  the $\sim 40 Hz$ scale  of the classical  oscillations, but on
the much longer time scale of  the {\it difference} of the periods of  the two coupled modes,
there will be, in this model, by virtue  of the quantum mechanical 
effects associated with a rapid sequence of repeated probing actions, 
a strong tendency for the brain correlate of consciousness to follow 
the {\it expected} trajectory, in contrast to what would happen if 
only infrequent probing actions were made.

This analysis is based on a theory of the mind-brain connection that 
resolves in principle the basic interpretational problem of quantum 
theory, which is the problem of reconciling the classical character 
of our perceptions of the physical world with the non-classical 
character of the state of the world generated by the combination of 
the Schroedinger equation and the uncertainty principle. The theory 
resolves also the central problem of the philosophy of mind, which is 
to reconcile the apparent causal power of our conscious efforts
with the laws and principles of physics. This relatively simple theory 
allows us to understand within the {\it dynamical framework}  
of orthodox (knowledge-associated-collapse) quantum physics the evident
capacity of our conscious  thoughts to influence our physical actions,
and to become thereby integrated into the process of natural selection.

The discussion  has focused  so far  on one very  small region  of the
brain, or rather on one small region together with an environment into
which it would, in the absence of probing observations, 
dissipate its energy. But the possible experiences of the relevant kind are 
associated with synchronous excitations in a large collection of such 
localized regions\cite{fell,engel}. Following the 
principles of quantum {\it field} theory the associated 
quantum state is represented by a {\it tensor product} of states 
associated with the individual  tiny regions.  

There has been a lot of detailed theoretical work examining the effects of the
fact that for a system that extends over an appreciable region of spacetime,
the parts of the systen located in different regions are coupled effectively to
different degrees of freedom of the environment\cite{decoherence}. Insofar as 
these aspects of the environment are never observed, the predictions of quantum
theory are correspondingly curtailed. In particular, {\it relative phases} of the wave
function of the system associated with different regions becomes impossible to 
determine, and a ``superposition'' of spatially separated components
becomes reduced to a ``mixture''.

In the model under consideration here the components in different space-time regions
are different {\it factors} of a tensor product, rather than different terms
of a superposition. In this case, the fact that different regions of the system
are coupled to different degrees of freedom of the environment does not
produce the usual quantum decoherence effects.

\section{Acknowledgements}

I thank Efstratios Manousakis, Kathryn Laskey, Edward Kelly, Tim Eastman,
Ken Augustyn, and Stan Klein for valuable suggestions.

This work was supported by  the Director, Office of Science, Office of
High  Energy and  Nuclear Physics,  of the  U.S. Department  of Energy
under contract DE-AC02-05CH11231

\end{document}